\documentclass[]{revtex4}
\usepackage{graphicx}
\usepackage{latexsym}
\usepackage{amssymb}

\newcommand {\eq}{\begin{equation}}
\newcommand {\ee}{\end{equation}}

\begin{document}

\title{Observational Constraints on Agegraphic Dark Energy}

\author{ Xing Wu$^{1}\footnote{Email address: wxxwwxxw@mail.bnu.edu.cn},
 $Yi Zhang$^{2},$ Hui Li$^{2}, $ Rong-Gen Cai$^{2},$ and Zong-Hong Zhu$^{1}$}
\address{$^{1}$Department of Astronomy, Beijing Normal University,
Beijing 100875, China}
\address{$^{2}$ Institute of Theoretical
Physics, Chinese Academy of Sciences, P.O. Box 2735, Beijing 100080,
China}

\begin{abstract}
In this paper, we use the Type Ia supernova data as well as the CMB
and LSS data to constrain the agegraphic dark energy model recently
proposed by Cai. Due to its peculiar nature, the parameter $n$ of
this model cannot be well constrained by the SNIa data, while the
other parameter $\Omega_{m0}$ can be constrained to be
$0.34\pm0.04$. When combined with CMB and LSS data, the range of
$1\sigma$ confidence level for $n$ is greatly narrowed, albeit still
very large. The best fit result is $\Omega_{m0}=0.28\pm0.02$, which
is consistent with most observations like WMAP and SDSS, and
$n=3.4$, of which a meaningful range of confidence level can not be
obtained due to the fact that the contours are not closed. Despite
of this result, we conclude that for $n>1$ this model is consistent
with SNIa, CMB and LSS observations. Furthermore, the fitting
results indicate a generalized definition for the agegraphic dark
energy.

\end{abstract}


\maketitle

\section{Introduction}
The dark energy problem has become undoubtedly one of the most
challenging  problems in modern physics ever since the discovery of
the accelerating expansion of the universe\cite{acc}. Various
theoretical models to explain the origin of the cosmic acceleration
have been proposed (for a recent review, see \cite{0603057} and the
references therein). Although the cosmological constant is the
simplest one among other models, it suffers from the so-called
coincidence problem and the parameter in this model has to be
fine-tuned in order to be consistent with observations\cite{cc}. One
of the reasons for this awkwardness is that it identifies the
cosmological constant with the vacuum energy of the quantum field
theory in Minkowski spacetime. At the cosmological scales, however,
the effect of gravity is significant and the above result may break
down. A complete solution to the cosmological problem is expected to
be given by a full theory of quantum gravity, which is unknown as
yet. Luckily, the so-called Holographic Principle \cite{holo.
principle} has shown some important features of quantum gravity.
Based on this principle, Cohen et. al.\cite{cohen} suggested a
relation between the IR cut-off and the UV cut-off in quantum field
theory. That is, in a box of size $L$, the quantum zero-point
energy, related to the UV cut-off, should not exceed the mass of a
black hole of the same size. This leads to the holographic vacuum
energy given by \eq \rho_{\Lambda}=\frac{3c^2M_p^2}{ L^2}~,\ee where
$M_p\equiv(8\pi G)^{-1/2}$ is the reduced Planck mass and $3c^2$ is
by convention a numerical factor. Although $L$ chosen as the Hubble
scale $H_0^{-1}$ or the particle horizon can lead to an energy
density consistent with current observation, the equation of state
for the vacuum energy in these two cases is always larger
than$-1/3$\cite{Hsu}\cite{Li}, therefore they cannot play the role
of dark energy. Li\cite{Li} proposed to choose $L$ as the future
event horizon leading to the model of holographic dark energy which
gives a correct equation of state. This model has been tested by
observational data\cite{holotest1}-\cite{holotest5} and is
consistent with them. However, it is plagued on the fundamental
level due to its assumption that the current properties of the dark
energy is determined by the future evolution of the universe, which
seems to violate causality. Moreover, it has been argued that this
model can be inconsistent with the age of the universe\cite{wh}.

Recently, Cai\cite{cai} proposed a model, called 'agegraphic dark
energy', based on the K\'{a}rolyh\'{a}zy uncertainty relation\cite{K
uncertain} \eq \delta t=\beta t_p^{2/3}t^{1/3}~, \ee where $\beta$
is a numerical factor of order one and $t_p$ is the Planck time.
This relation arises from quantum mechanics combined with general
relativity, and it imposes an upper limit of accuracy in any
measurement of distance $t$ (with the speed of light $c=1$) in
Minkowski spacetime. Combined with the time-energy uncertainty
relation in quantum mechanics, the energy density of the metric
fluctuations in Minkowski spacetime is\cite{11,12} \eq \rho_q\sim
\frac{E_{\delta t^3}}{\delta
t^3}\sim\frac{1}{t_p^2t^2}\label{fluctuation}~,\ee where $E_{\delta
t^3}$ is the energy fluctuation within a cell of size $\delta t$.
Based on this relation, Cai proposed the agegraphic dark energy
as\cite{cai}
 \eq \rho_q=\frac{3n^2M_p^2}{T^2}\label{rho}~, \ee where $T$ is
the age of the universe defined as \eq T=\int_0^t
dt'=\int_0^a\frac{da}{Ha}=\int_z^\infty \frac{dz}{(1+z)H}\label{T
def}~,\ee and $n$ is a parameter of order one representing some
uncertainties, such as the species of quantum fields in the universe
or the effect of curved spacetime (since equation
(\ref{fluctuation}) is derived for Minkowski spacetime). The form of
this model mimics a model of holographic dark energy with $L$ chosen
as the age of the universe. It is not surprising since, as is
mentioned in\cite{cai}, the K\'{a}rolyh\'{a}zy uncertainty relation
is also a reflection of the entanglement between UV and IR scale in
effective quantum field theory. Moreover, due to using the age of
the universe instead of the future event horizon, the new model is
free of the causality problem which undermines the holographic dark
energy model. The effect of the interaction between the agegraphic
dark energy and dark matter on the cosmological evolution has been
investigated in \cite{WeiCai1,WeiCai2}

This paper is organized as follows. In Sec.~II we introduce the
agegraphic dark energy model and compare its features with those of
the holographic dark energy model. In Sec.~III we use data from Type
Ia supernova as well as CMB and LSS observations to constrain the
parameters of the model. Sec.~IV is devoted to conclusions.

\section{The dark energy model}
In this section we review the agegraphic dark energy model proposed
in \cite{cai}. Then we confront this model with the Type Ia
supernova observation as well as a joint analysis together with CMB
and large scale structure(LSS) data in the next section. Now we
assume a spatially flat FRW universe with matter component $\rho_m$
and the agegraphic dark energy $\rho_q$. The Friedmann equation is
\eq H^2=\frac{1}{3M_p^2}(\rho_q+\rho_m)\label{Friedmann}~,\ee where
$\rho_m=\rho_{m0}(1+z)^3$. We define $\Omega_q=\rho_q/3M_p^2H^2$ and
$\Omega_m=\rho_m/3M_p^2H_0^2=\Omega_{m0}(1+z)^3$ and recast equation
(\ref{Friedmann}) into \eq
E=H/H_0=\sqrt{\frac{\Omega_m}{1-\Omega_q}} \label{E}~.\ee Once the
evolution of $\Omega_q$ is known, $E$ is determined. To find out how
$\Omega_q$ evolves with time(redshift), we combine equation
(\ref{rho}) and (\ref{T def}) to get \eq
\int_z^{\infty}\frac{dz}{(1+z)H}=\frac{n}{H\sqrt{\Omega_q}}\label{T}~.\ee
Taking derivative with respect to $z$ in both sides of equation
(\ref{T}) leads to \eq
\Omega'_q=\frac{-1}{1+z}(3-\frac{2}{n}\sqrt{\Omega_q})\Omega_q(1-\Omega_q)
\label{domega_q}~,\ee where the prime denotes $d/dz$. With the
initial condition implied by setting $z=0$ in equation
(\ref{Friedmann})\eq \Omega_{q0}+\Omega_{m0}=1 \label{initial}~,\ee
we can solve this differential equation to obtain the evolution of
$\Omega_q$, and finally determine $E(z)$. Here one point is worth
particular mentioning. As mentioned in \cite{cai}, adding a
z-independent term $\delta_q$ to the LHS of equation (\ref{T}) leads
to the same differential equation (\ref{domega_q}), and moreover,
imposing the initial condition (\ref{initial}) cannot guarantee a
vanishing $\delta_q$\footnote{It is easy to see that the same
situation also exists in the holographic dark energy model.}. As a
result, the solution to equation (\ref{domega_q}) leads to a dark
energy density which should be written in a more general form as \eq
\rho_q=\frac{3n^2M_p^2}{(\delta_q+T)^2}\label{rho general}~, \ee
where $\delta_q$ may generally be a function of $n$ as well as
$\Omega_{m0}$ and $H_0$, which can be calculated by \eq
\delta_q={n\over{H(z)\sqrt{\Omega_q(z)}}}-
\int_z^{\infty}\frac{dz'}{(1+z')H(z')} \label{delta_q}.\ee Note that
although $z$ explicitly emerges in the RHS of the above expression,
$\delta_q$ is independent of $z$ due to the subtraction of the two
terms.

\begin{figure}[htbp]
\begin{center}
\includegraphics[scale=0.50]{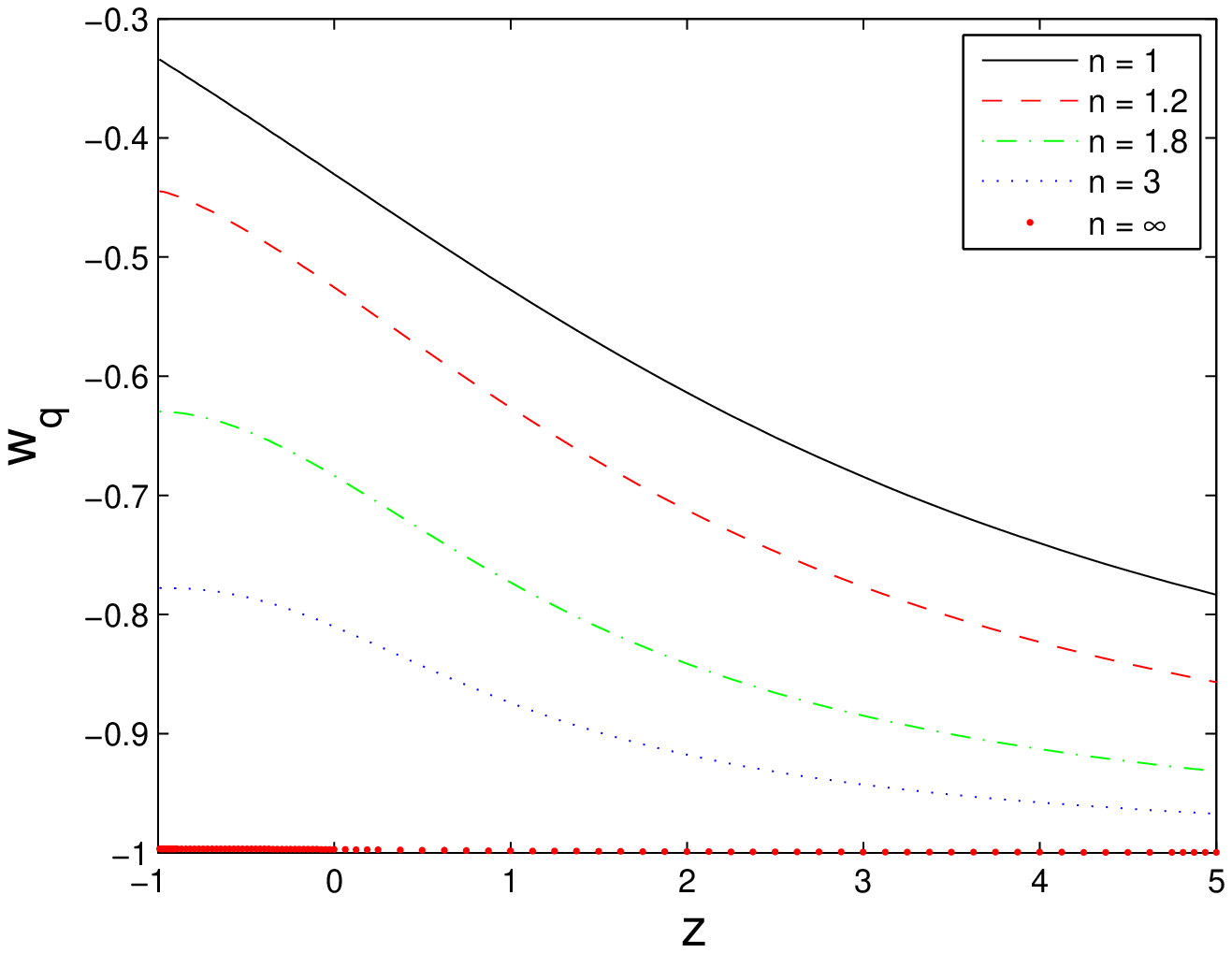}   \quad
\includegraphics[scale=0.50]{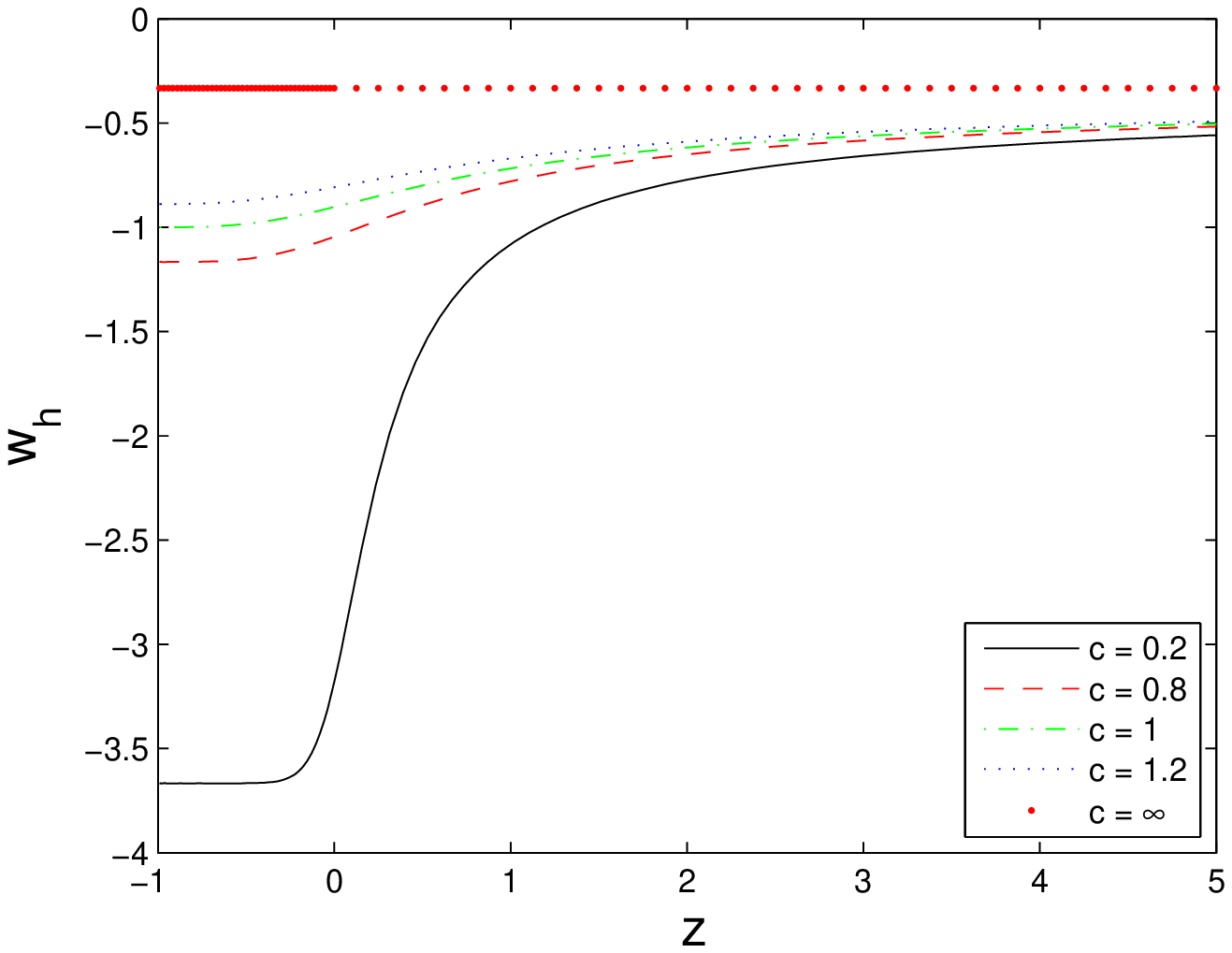}
\caption{The evolution of $w_q$ (left) and $w_h$ (right) with
respect to redshift $z$ with varying parameters $n$ and $c$. The red
dotted lines denote the asymptotic behavior as $c$ or $n$ increases.
} \label{fig:w}
\end{center}
\end{figure}

Then we consider the equation of state of the dark energy. From the
conservation equation \eq \dot{\rho_q}+3H(1+w_q)\rho_q=0~,\ee we
derive \eq w_q=-1-\frac{\dot{\rho_q}}{3H\rho_q}~, \ee which leads to
\eq w_q=-1+\frac{2}{3n}\sqrt{\Omega_q}~.\ee The quantum fluctuations
can serve as dark energy for $w_q<-1/3$, that is,
$n>\sqrt{\Omega_q}$. For example, if we take $\Omega_{q0}=0.73$, as
indicated by WMAP\cite{WMAP} for LCDM, $n_0$ should be larger than
$0.85$ to behave as the dark energy. In fact, in order to drive the
cosmic acceleration, $n_0$ should be even larger given the fact that
currently the matter component is not negligible and the total
effective equation of state should be taken into account. In
addition, it is shown in \cite{cai} that $\rho_q$ becomes dominant
in the future, namely $\Omega_q\rightarrow 1$. Therefore, the
universe will eternally accelerate if $n>1$; for $n<1$, current
cosmic acceleration would be a transient process and the expansion
would finally slow down in the future. We note that the same
phenomenon may appear in the branewrold scenario as discussed
in\cite{sahni}. The evolution of $w_q$ is shown in figure
\ref{fig:w}.
\begin{figure}[htbp]
\begin{center}

\includegraphics[scale=0.50]{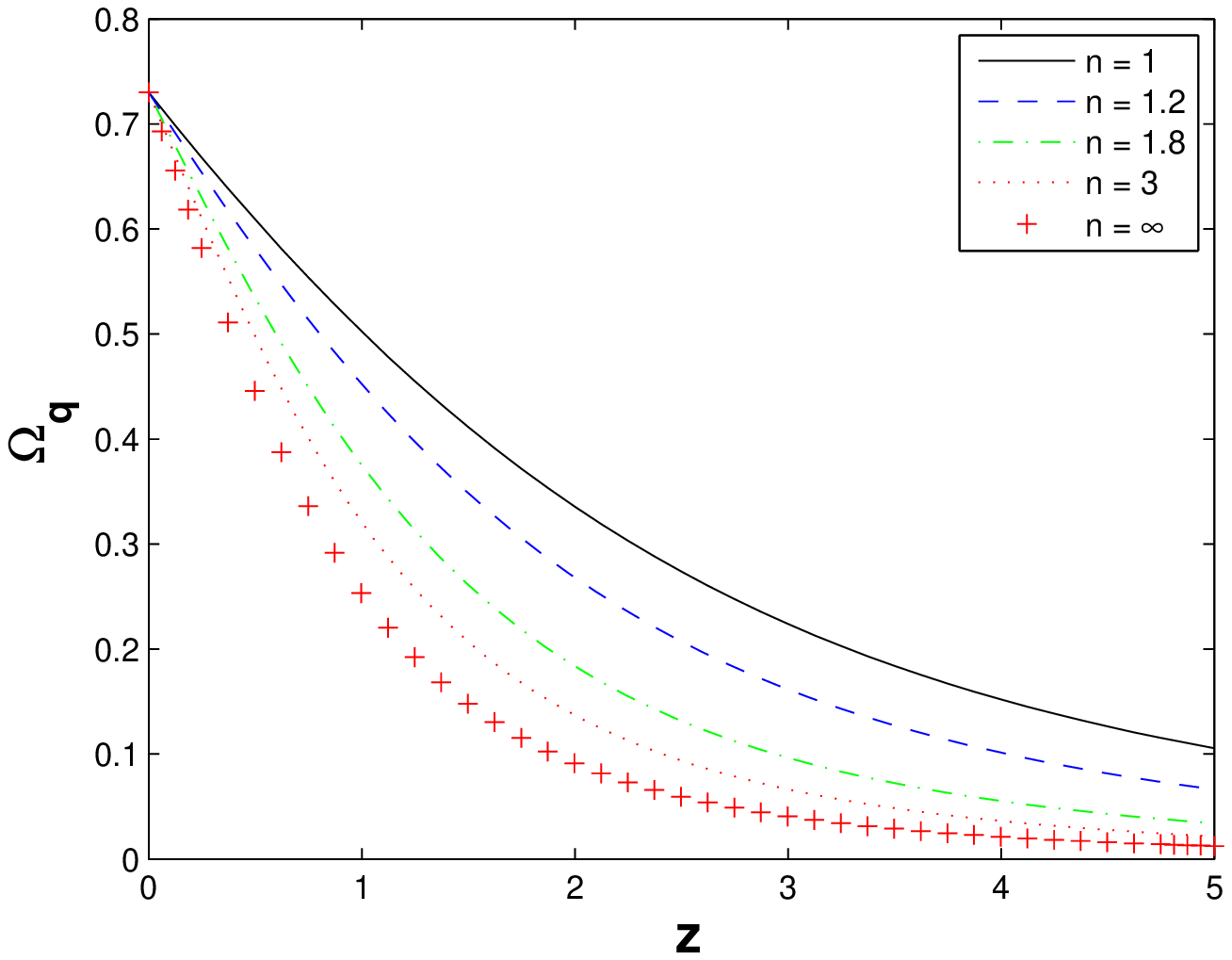}   \quad
\includegraphics[scale=0.50]{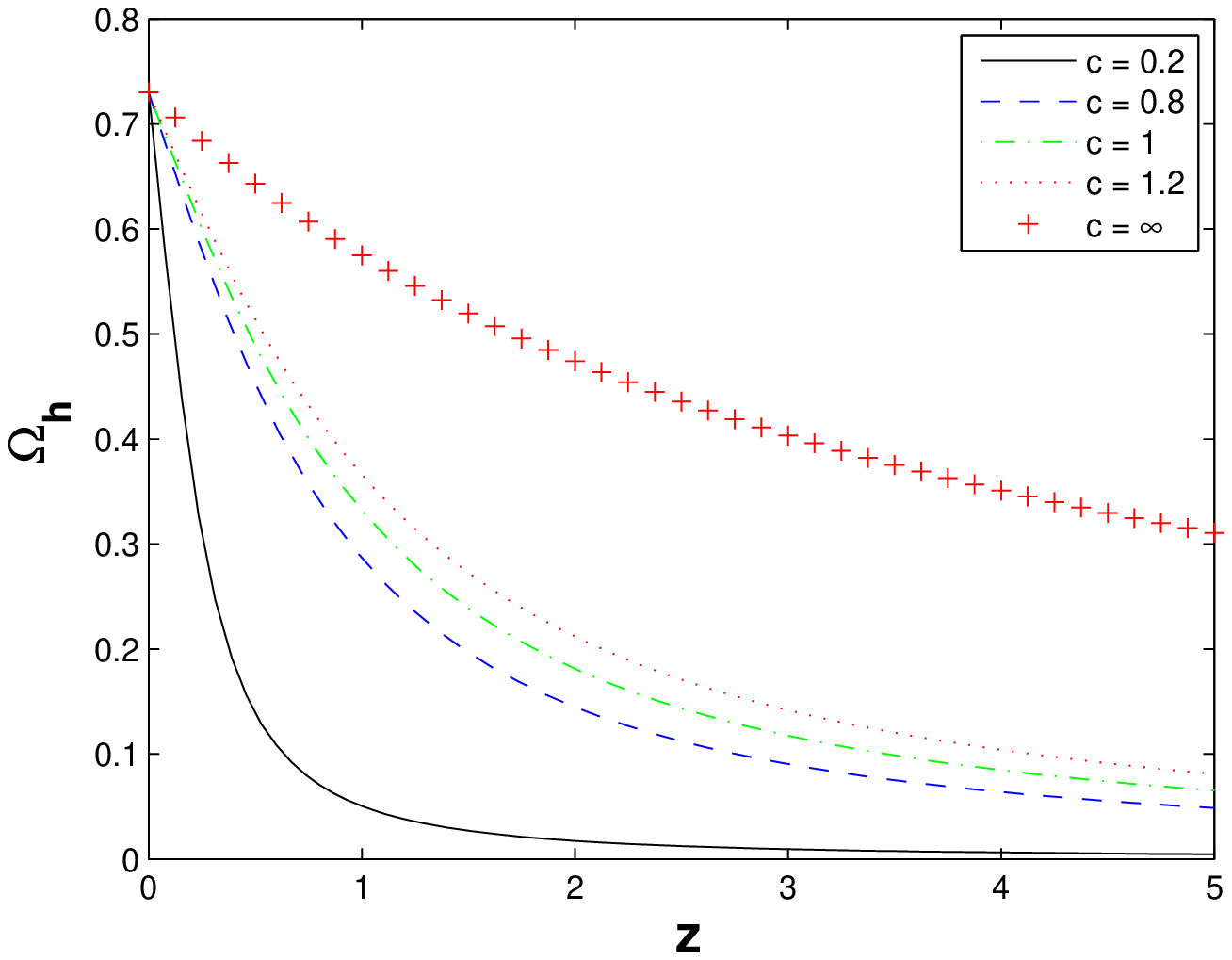}
\caption{The evolution of $\Omega_q$ (left) and $\Omega_h$ (right)
with respect to redshift $z$ with varying parameters $n$ and $c$.
Here we assume $\Omega_{m0}=0.27$. For $\Omega_q$ the traces
asymptotically tend to a fixed curve (denoted by the red $+$'s) as a
lower bound when $n$ increases; for $\Omega_h$ the traces
asymptotically tend to a fixed curve (denoted by the red $+$'s) as
an upper bound when $c$ increases. Of course the values of the
parameter much lager than 1 is not physical, here we just use them
as an illustration for the asymptotic behavior of the corresponding
differential equations. } \label{fig:omega}
\end{center}
\end{figure}
It is interesting to compare this model with the holographic dark
energy, for which we use a subscript 'h' to denote its corresponding
quantities. For $\Omega_h$, its evolution is determined by \eq
\Omega'_h=\frac{-1}{1+z}\underline{(1+\frac{2}{c}\sqrt{\Omega_h})}\Omega_h(1-\Omega_h)
\label{domega_h}~,\ee with the same initial condition as
(\ref{initial}). It is the underlined part that indicates the
difference in the two models (if we identify $n$ with $c$).  For
illustration, the evolution curves for both $\Omega_q$ and
$\Omega_h$ with respect to $z$ are plotted in figure
\ref{fig:omega}. Note that equation (\ref{domega_q}) and
(\ref{domega_h}) only depend on one parameter $n$ and $c$
respectively. The figure shows that for $\Omega_q$, the curve moves
downwards as $n$ increases and asymptotically tends to a fixed
position as a lower bound; for $\Omega_h$, the curve moves upward as
$c$ increases and ends up asymptotically at a fixed position as an
upper bound. The asymptotic behavior of the fractional energy
density for large $c$ or $n$ in the two models can easily be seen
from the corresponding differential equations: \eq
\Omega'_q=\frac{-1}{1+z}3\Omega_q(1-\Omega_q)\label{domega_limit}~,\ee
and
 \eq
\Omega'_h=\frac{-1}{1+z}\Omega_h(1-\Omega_h)~.\ee

 As for the equation of state, for holographic dark energy \eq
w_h=-\frac{1}{3}-\frac{2}{3c}\sqrt{\Omega_h}~.\ee Considering its
future evolution, for $c\geq1$,$w_h>-1$ forever, while for $c<1$,
$w_h$ may cross the phantom divide and end up with $w_h<-1$. This is
a significant difference from the agegraphic dark energy, where
$w_q$ can never be less than $-1$. In addition, as $z$ grows,
$w_h\rightarrow -1/3$ whereas $w_q\rightarrow -1$ implying the
agegraphic dark energy behaves like a cosmological constant at early
time. This difference in high redshift region may exert different
influence on the evolution of the early universe, such as the matter
fluctuation and the formation of large scale structure. Further
analysis based on cosmological perturbation theory together with the
CMB observation may shed some light on distinguishing these two
models. The evolution of the equation of state for both models is
plotted in figure  \ref{fig:w} for comparison.

\section{Constraints from Type Ia SNe observation}
\begin{figure}[htbp]
\begin{center}
\includegraphics[scale=0.7]{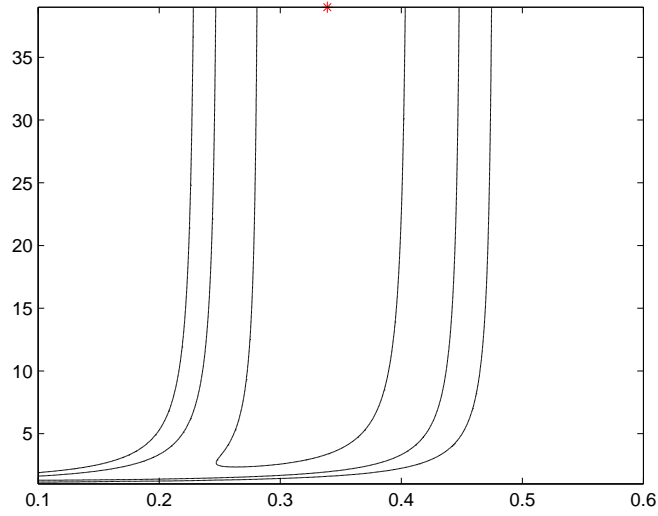}   \quad
\caption{$68.3\%$, $95.4\%$ and $99\%$ confidence contours for the
parameters $(n,\Omega_{m0})$. It is obvious that the range for $n<1$
is outside the $3\sigma$ confidence contour. } \label{fig:chi2_SN}
\end{center}
\end{figure}

\begin{figure}[htbp]
\begin{center}

\includegraphics[scale=.7]{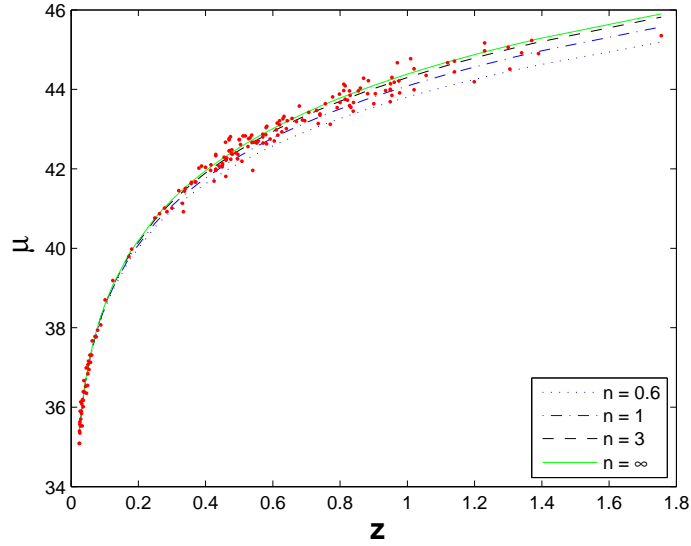}   \quad
\caption{Illustration for the weakness of using SN data to constrain
the agegraphic dark energy model. Here we assume $\Omega_{m0}=0.28$
and $h=0.62$} \label{fig:Hubble}
\end{center}
\end{figure}

Now we perform the $\chi^2$ statistics to constrain the parameters
$(n,\Omega_{m0})$ of the model in question with the goldset of 182
SNIa data compiled by Riess et.al.\cite{gold}. The observations of
supernovae measure the apparent magnitude $m$, which is related to
the luminous distance $d_L$ of an object (SN) at redshift $z$ as \eq
m(z)=M+5\log d_L(z)+25~, \ee where $M$ is the absolute magnitude,
which can generally be considered to be the same for Type Ia
supernovae as the standard candles. In a flat universe \eq
d_L=H_0^{-1}(1+z)\int_0^z\frac{dz'}{E(z')}\label{dl}~,\ee where the
Hubble scale $H^{-1}_0=2997.9h^{-1}Mpc$. In order to constrain the
parameters, we compute the distance modulus $\mu=m-M$, and minimize
the quantity $\chi^2$ defined by \eq
\chi^2(n,\Omega_{m0})=\sum_i^{182}
\frac{[\mu_{obs}(z_i)-\mu_{th}(z_i;n,\Omega_{m0})]^2}{\sigma_i^2}~,\ee
where $\sigma_i$ is the observational uncertainty. Assuming that the
errors are Gaussian, the likelihood is $\mathcal{L}\propto
e^{-\chi^2/2}$. In figure \ref{fig:chi2_SN}, we plot the contours of
confidence level at $68.3\%$, $95.4\%$ and $99\%$ in the
$(n,\Omega_{m0})$ plane with $h$ marginalized. The best fit values
corresponding to the minimum of $\chi^2$ are $n=39$ and
$\Omega_{m0}=0.34$, which are denoted by a red star on the plot. As
we see, the parameter $\Omega_{m0}$ is well constrained, and the
range $n<1$ is outside the $3\sigma$ confidence level. This
indicates that an accelerated expansion in our model is compatible
with current observations. However, the contours are not closed from
above within the given range , and the point of the best fit value
is on the border of the upper limit of $1<n<39$. It seems that the
range of $n$ is not large enough to encompass the best fit value. We
point out that it is the particularity of this model that makes the
parameter $n$ cannot be well constrained by SN data. As an
illustration, we use equation (\ref{domega_limit}) to calculate the
$\chi^2_{min}$ for $n\rightarrow\infty$. The result is
$\chi^2_{min}|_{n=\infty}=158.7422$ with the best fit
$\Omega_{m0}=0.34$. Comparing this with $\chi^2_{min}=158.8526$
within the range $1<n<39$, the difference is negligible given the
fact that $n$ spans such a large range from $39$ to $\infty$. This
implies that the parameter $n$ within such a large range is always
consistent with SN data. We further illustrate this in figure
\ref{fig:Hubble}. As we see, in the low redshift region, the curves
corresponding to different values of $n$ are indistinguishable. For
high redshift, however, we see that for $n<1$ (e.g. $n=0.6$ in the
plot) the curve predicted by the model is remarkably not consistent
with the data points, whereas for $n>1$, all the curves, including
the curve for $n=\infty$(the green line), are confined within a very
narrow bunch which seems not significantly inconsistent with the
data. Therefore, this model cannot be well constrained by using SN
data alone.


This seems surprising since almost all models of dark energy can be
constrained significantly by SN data. There would be no surprise,
however, once we recall the asymptotic behavior of $\Omega_q$ with
increasing $n$. In figure \ref{fig:omega} it shows that the larger
$n$ is, the less difference it makes between two neighboring curves.
Therefore, as $n$ grows, changes in the quantity $\chi^2$ become
less significant. In fact there is also a similar asymptotic
behavior in the holographic dark energy, where the $\Omega_h$
becomes insensitive to the change of $n$ as $n$ grows large. In that
case, the best fit value $c=0.2 \sim 0.9$
\cite{holotest1}-\cite{holotest5} lies in the range where $\Omega_h$
is sensitive to varying $n$. Thus the holographic model can be well
constrained by SN data. Moreover, figure \ref{fig:omega} also shows
that the range where $\Omega_h$ is sensitive to $c$ is just that
where $\Omega_q$ is insensitive to $n$, and vice versa. Thus, if we
assume that the evolution of dark energy in both models should not
be different significantly in the same range, we find that
unfortunately the possible 'best fit' value for $n$ falls in the
range of insensitivity.

\begin{figure}[tbhp]
\begin{center}
\includegraphics[scale=0.7]{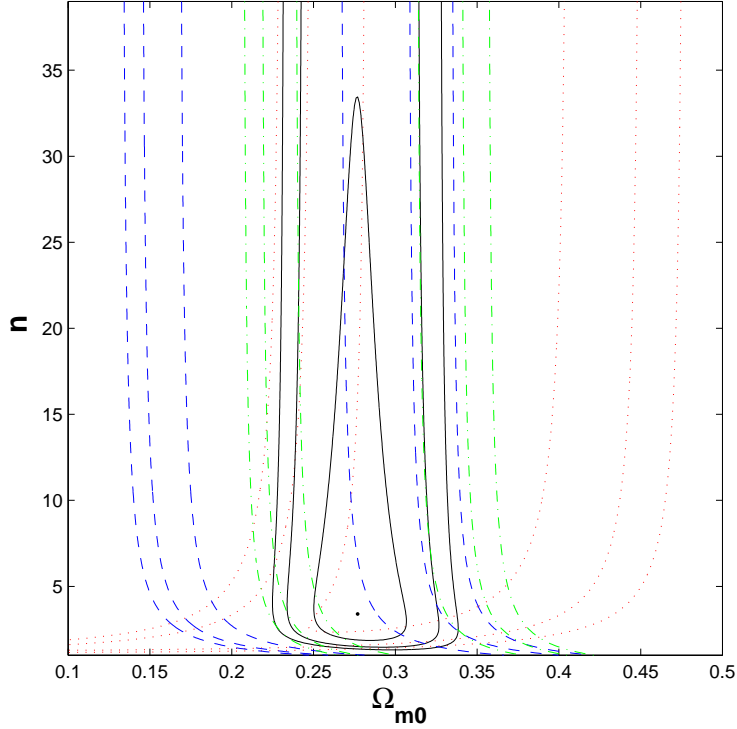}   \quad

\caption{Confidence contour plot of $68.3\%$, $95.4\%$ and $99\%$
with the red dotted line for SN data, the blue dashed line for CMB,
the green dash dotted line for BAO, and the black line for the joint
analysis. We marginalized the nuisance parameter $h$. The best fit
values are $\Omega_{m0}=0.28$,$n=3.4$} \label{fig:contour_com}
\end{center}
\end{figure}

In order to improve the situation, we combine the constraints from
CMB\cite{WMAP} and LSS\cite{SDSS} observations. The CMB shift
parameter is the most model-independent parameter we can extract
from the CMB data, which is given by\cite{R} \eq
R=\sqrt{\Omega_{m0}}\int^{z_{rec}}_0\frac{dz}{E(z)}\label{R}~, \ee
where $z_{rec}=1089$ represents the redshift of recombination. This
parameter relates the angular diameter distance to the last
scattering surface, the comoving size of the sound horizon at
$z_{rec}$ and the angular scale of the first acoustic peak in CMB
power spectrum of temperature fluctuations. Here we assume the value
$R=1.70\pm0.03$ given by Wang and Mukherjee\cite{R value}. The
$\chi^2$ value is \eq \chi^2_{CMB}=\frac{(R-1.70)^2}{0.03^2}~.\ee
For LSS data, we use the measurement of the baryon acoustic peak
(BAO) in the distribution of SDSS luminous red galaxies
(LRG's)\cite{SDSS}. This peak is the imprint left by the cosmic
perturbation in early universe on the late-time non-relativistic
matters, and it can provide an independent constraint on dark energy
models. We use here the parameter $A$ defined as \eq
A=\sqrt{\Omega_{m0}}E(z_1)^{-1/3}\left
(\frac{1}{z_1}\int_0^{z_1}\frac{dz}{E(z)}\right)^{2/3}\label{A}~,\ee
where $z_1=0.35$ is the typical LGR redshift. The observational
value is given by Eisenstein et al\cite{BAO} as $A=0.469\pm0.017$.
And the $\chi^2$ value is \eq
\chi^2_{BAO}=\frac{(A-0.469)^2}{0.017^2}~.\ee Now we perform a joint
analysis by minimizing the combined quantity
\eq\chi^2=\chi^2_{SN}+\chi^2_{CMB}+\chi^2_{BAO}~.\ee The result is
plotted in figure \ref{fig:contour_com}, with the red dotted line
for SN data, the blue dashed line for the shift parameter R, the
green dash dotted line for parameter A and the black solid line for
the combined contours. We can see the contour of $1\sigma$ is
closed, although the range is still large. The best fit values are
$\Omega_{m0}=0.28$ and $n=3.4$. Accordingly, the current equation of
state is $w_{q0}=-0.83$, which is consistent with the WMAP
observation\cite{WMAP}.

\begin{figure}[htbp]
\begin{center}
\includegraphics[scale=0.50]{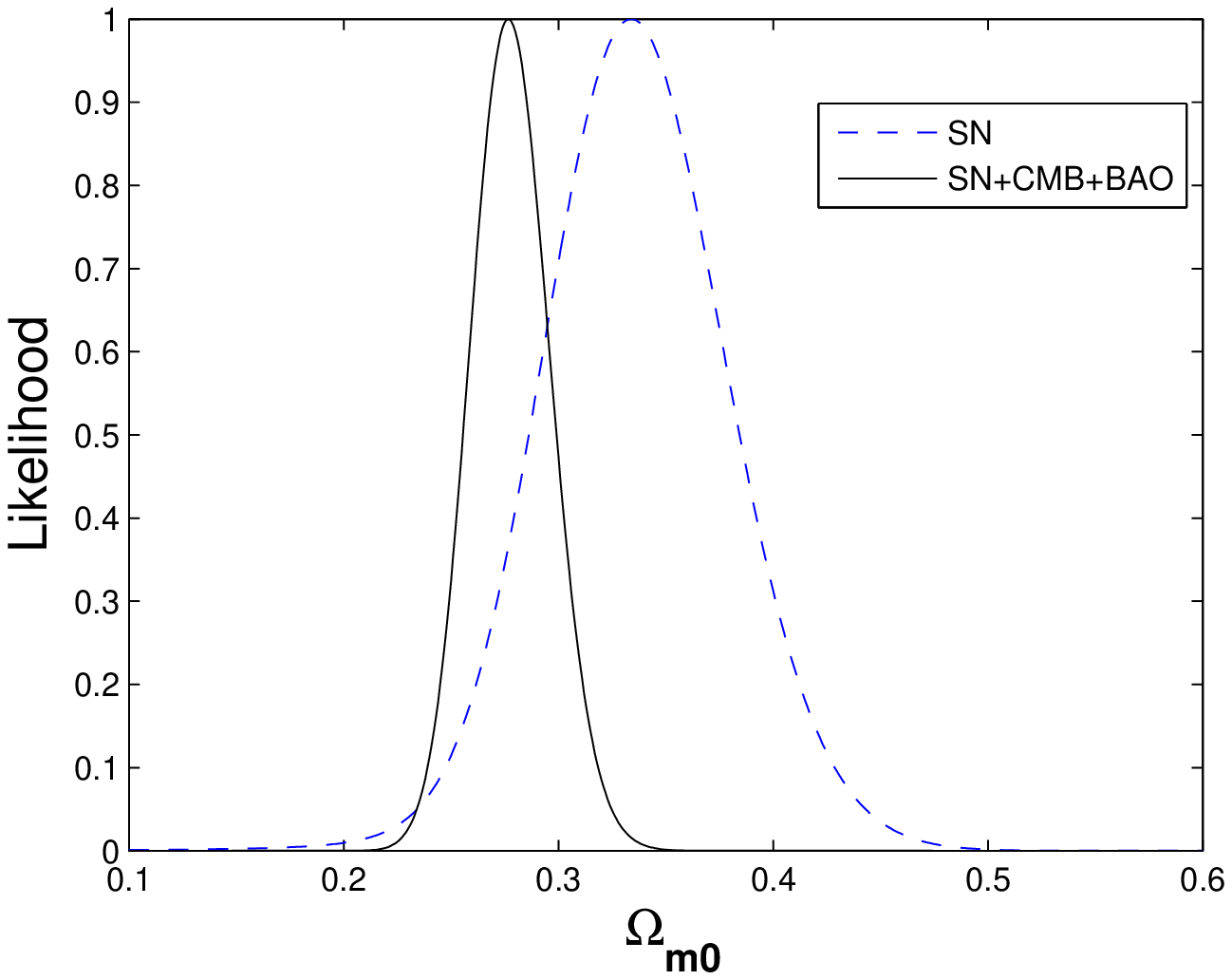}   \quad
\includegraphics[scale=0.50]{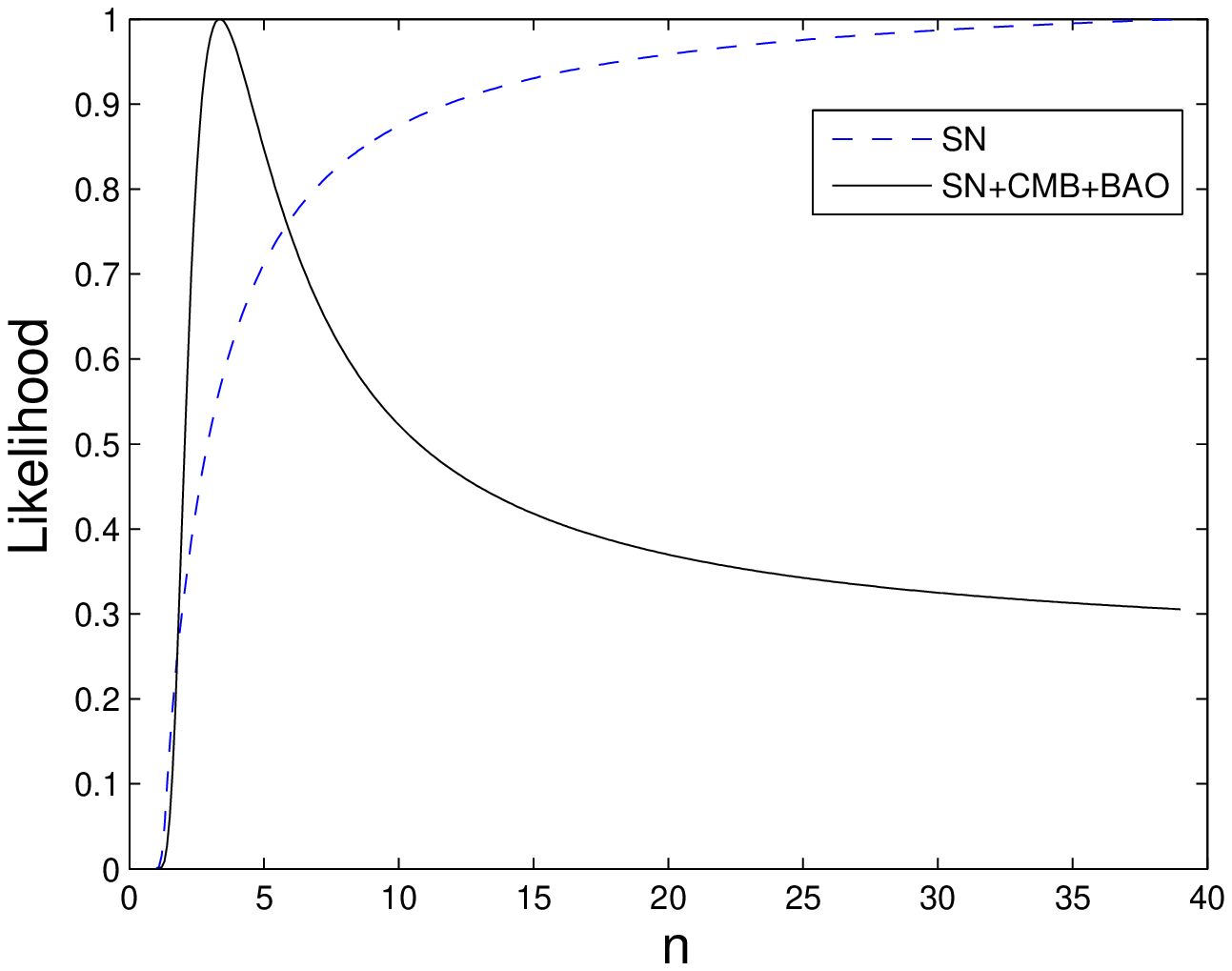}
\caption{The likelihood for $\Omega_{m0}$ and $n$. The black line
corresponds to the joint analysis, and the blue dotted line
corresponds to using only SN data. Note that for the likelihood
function for $n$, the plot asymptotically tends to a horizontal line
as $n\rightarrow\infty$, which means the area under the plot is
divergent. This actually implies that $n$ cannot be well constrained
by these observations.} \label{fig:likelihood}
\end{center}
\end{figure}

The likelihood functions for $\Omega_{m0}$ and $n$ are shown in
figure \ref{fig:likelihood}. As we can see, the $\Omega_{m0}$
likelihood has a near-gaussian shape, therefore we can give the best
fit at $1\sigma$ by integrating to $68.3\%$ of the total area under
the curve. The results are $\Omega_{m0}=0.33\pm0.04$ for using SN
data alone and $\Omega_{m0}=0.28\pm0.02$ for the joint analysis. For
$n$, however, the likelihood curve is highly asymmetric and, what is
more, the total area is divergent. Thus we cannot extract a range
for the best fit $n=3.4$ with definite statistic meaning. As a
consequence, although we use the terms such as $68.3\%$, $95.4\%$
and $99\%$ confidence level, they are not valid in a strict sense.
Here we just use them by convention to denote the contours
corresponding to $\Delta\chi^2=\chi^2-\chi^2_{min}=2.30,6.17$ and
$9.21$ respectively.

\section{Conclusion and Discussion}
In this paper, we briefly reviewed the agegraphic dark energy model
based on the K\'{a}rolyh\'{a}zy uncertainty relation, and then
compared it with the holographic dark energy, which is motivated by
the holographic principle. In fact both models based on some
principles that relate the IR and UV cut-offs in effective quantum
field theory, therefore they have similar features. Apart from the
essential distinction in physical motivations, the apparent
difference lies in that for the agegraphic dark energy the IR
cut-off is chosen as the age of the universe and the equation of
state $w_q>-1$ forever, whereas for the holographic dark energy the
length measure is the future event horizon and $w_h$ can cross $-1$.
When using the Type Ia supernova data to constrain the model, we
find that although $\Omega_{m0}$ can be well constrained, the
parameter $n$ is unbounded from above. After a joint analysis
together with the CMB shift parameter $R$ and the BAO $A$ parameter,
$\Omega_{m0}$ is enhanced from $\Omega_{m0}=0.33\pm0.04$ to
$\Omega_{m0}=0.28\pm0.02$. The situation for $n$ is essentially
unchanged. As a result, this model can be consistent with current
SNIa data as well as the CMB and LSS data for $n>1$, which is just
the requirement of this model for an accelerated expansion, as is
mentioned in \cite{cai}.


\begin{figure}[htbp]
\begin{center}
\includegraphics[scale=0.50]{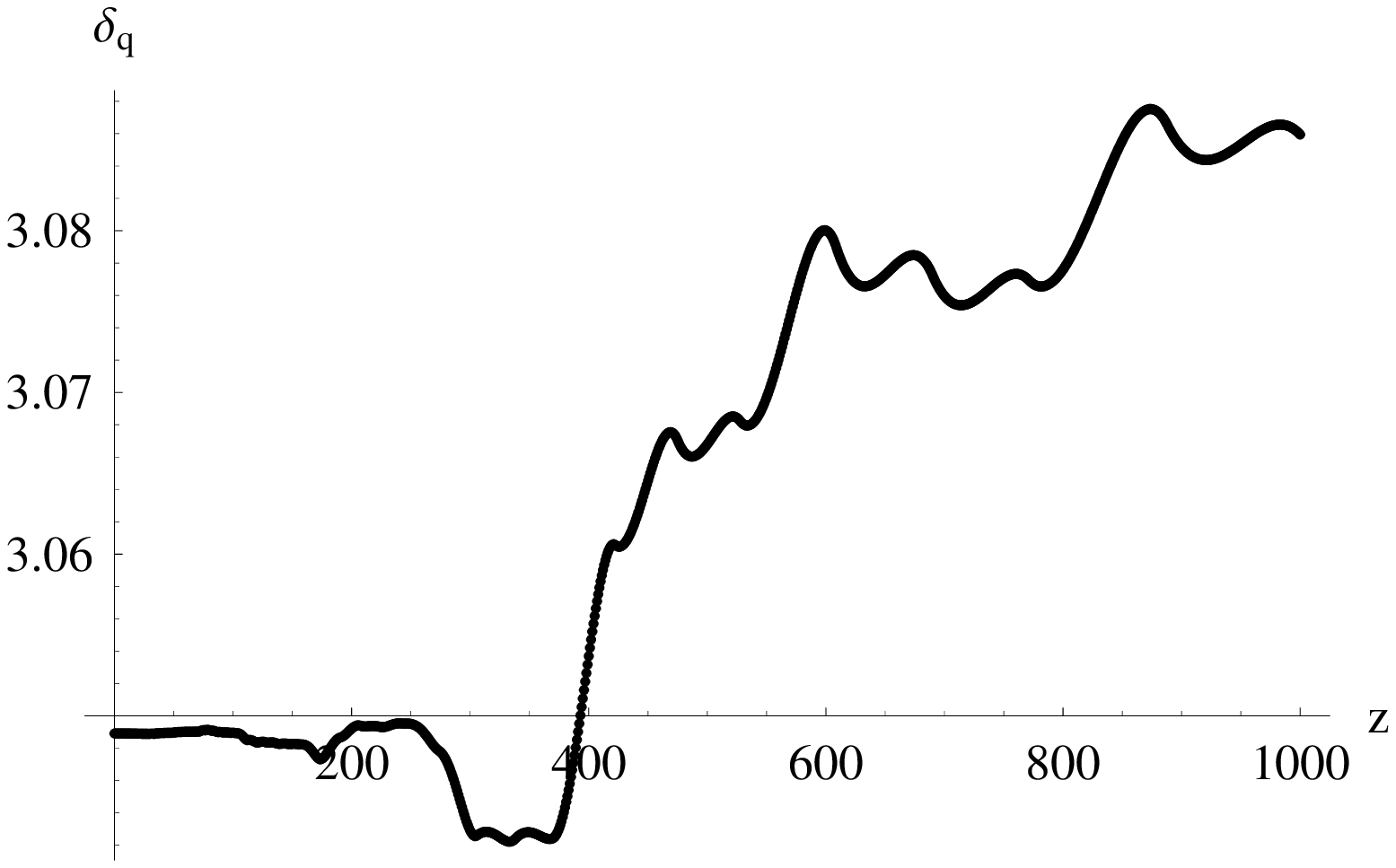}   \quad

\caption{For agegraphic dark energy, $\delta_q$ in unit of
$H_0^{-1}$ calculated with $z$ from the range $[0,1000]$.
$\Omega_{m0}=0.28$, $n=3.4$.} \label{fig:delta_q}
\end{center}
\end{figure}


\begin{figure}[htbp]
\begin{center}

\includegraphics[scale=0.50]{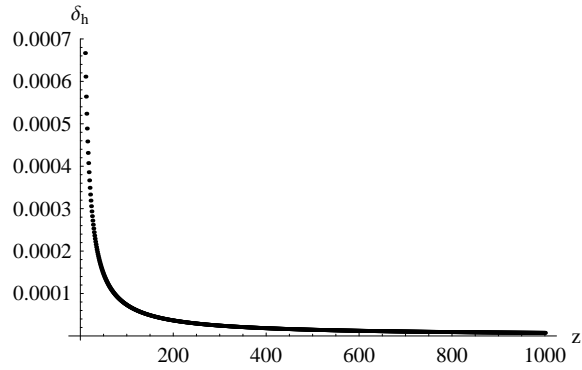}
\caption{For holographic dark energy, $\delta_h$ in unit of
$H_0^{-1}$ calculated with $z$ from the range $[0,1000]$.
$\Omega_{m0}=0.29$, $c=0.91$.} \label{fig:delta_h}
\end{center}
\end{figure}
As we mentioned before, the solution to equation (\ref{domega_q})
corresponds to the general definition for the agegraphic dark energy
(\ref{rho general}). With the best fits from the joined analysis, we
calculate $\delta_q$ by equation (\ref{delta_q}). By definition,
$\delta_q$ is independent of $z$, therefore choosing arbitrarily any
$z$ will lead to the same result. Here we choose the value from a
large range $0\leq z\leq1000$. From figure \ref{fig:delta_q} we may
safely draw the conclusion that $\delta_q\simeq 3.0$ (in unit of
$H_0^{-1}$) and the variation of $\delta$ is due to the errors from
numerical method. For comparison, we also calculate the
corresponding quantity in the holographic dark energy model, where
\eq \delta_h={c\over{H(z)\sqrt{\Omega_h(z)}}}-
{1\over(1+z)}\int_{-1}^z\frac{dz'}{H(z')} \label{delta_h}.\ee We
assume the best fits $\Omega_{m0}=0.29$ and $c=0.91$ from
\cite{holotest5}. Figure \ref{fig:delta_h} shows $\delta_h\lesssim
0.0007$. Such a tiny quantity may well be neglected in practical
consideration. But here, $\delta_q$ is of the same order as the
current age of the universe ($\sim {1\over H_0}$), and therefore has
to be taken into account. Such a term can be understood as a result
of generalizing the energy density of the quantum fluctuations in
Minkowski spacetime (\ref{fluctuation}) to the cosmological
scenario. More fundamental interpretation for $\delta_q$ and the
general definition of the agegraphic dark energy (\ref{rho general})
needs further studies.

\section*{Acknowledgments}
We are grateful to Hao Wei  for helpful discussions. XW would like
to thank Heng Yu for kind help.  XW and ZHZ were supported by the
National Natural Science Foundation of China, under Grant
No.10533010, 973 Program No.2007CB815401 and Program for New Century
Excellent Talents in University (NCET) of China. YZ, HL and RGC were
supported in part by a grant from Chinese Academy of Sciences (No.
KJCX3-SYW-N2), and by NSFC under grants No.~10325525, No.~10525060
and No.~90403029.

\end{document}